\documentclass[useAMS,usenatbib,usegraphicx]{mn2e}

\usepackage{aas_macros}

\title[The collapse of magnetized filaments]{Simulating the collapse of magnetized and turbulent filaments}
\title[The collapse of magnetized filaments]{The impact of magnetic fields and turbulence on the collapse of filaments}
\title[The collapse of magnetized filaments]{The impact of turbulence and magnetic field orientation on star forming filaments}
  \author[D. Seifried, S. Walch]
  {D.~Seifried,$^{1}$\thanks{seifried@ph1.uni-koeln.de} S.~Walch$^{1}$ \\
  $^1$I. Physikalisches Institut, Universit\"at zu K\"oln, Z\"ulpicher Str. 77, 50937 K\"oln, Germany}

\date{Released 2015}

\pagerange{\pageref{firstpage}--\pageref{lastpage}} \pubyear{2013}

\begin{document}

\label{firstpage}

\maketitle

\begin{abstract}
We present simulations of collapsing filaments studying the impact of turbulence and magnetic field morphologies on their evolution and star formation properties. We vary the mass per unit length of the filaments as well as the orientation of the magnetic field with respect to the major axis. We find that the filaments, which have no or a perpendicular magnetic field, typically reveal a smaller width than the universal width of 0.1 pc proposed by e.g. \citet{Arzoumanian11}. We show that this also holds in the presence of supersonic turbulence and that accretion driven turbulence is too weak to stabilize the filaments along their radial direction. On the other hand, we find that a magnetic field that is parallel to the major axis can stabilize the filament against radial collapse resulting in widths of 0.1 pc. Furthermore, depending on the filament mass and magnetic field configuration, gravitational collapse and fragmentation in filaments occurs either in an edge-on way, uniformly distributed across the entire length, or in a mixed way. In the presence of initially moderate density perturbations, a centralized collapse towards a common gravitational centre occurs. Our simulations can thus reproduce different modes of fragmentation observed recently in star forming filaments. Moreover, we find that turbulent motions influence the distance between individual fragments along the filament, which does not always match the results of a Jeans analysis.
\end{abstract}


\section{Introduction}

The importance of interstellar filaments for star formation has been highlighted for the first time by \citet{Schneider79}. More recently, the advent of the \textit{Herschel} satellite has uncovered the filamentary structure in molecular clouds and infrared dark clouds in great detail \citep[e.g.][but see also the review of \citealt{Andre13}]{Andre10,Konyves10,Arzoumanian11,Arzoumanian13,Peretto12,Schneider12,Palmeirim13}.

The \textit{Herschel} observations have revealed the interesting result that filaments apparently have a characteristic width of about 0.1 pc and a density profile, which can be represented by a Plummer-like profile. Furthermore, line emission observations of e.g. C$^{18}$O or N$_2$H$^+$ towards star forming filaments \citep{Arzoumanian13,Hacar13,Furuya14,Henshaw14,Jimenez14,Li14} report non-thermal line broadening, which points to an additional trans- to supersonic turbulent velocity field. Temperatures in filaments are typically in the range of $\sim$ 10 to 15 K, which is particularly important for calculating the critical mass per unit length of a filament \citep{Ostriker64}. This is the mass per length below which the filament can be stabilized against radial (i.e. perpendicular to the major axis) gravitational collapse. Filaments with a mass per unit length above the critical value are gravitationally unstable along the radial direction and will start to fragment along their major axis \citep[e.g][]{Myers09,Miettinen12,Kirk13,Kirk13b,Zernickel13}. Indeed, observations have shown that filaments with observed star formation activity typically have a mass per unit length above the critical one, whereas filaments with a mass per unit length below this value are mainly devoid of any star formation \citep{Andre10}.

Polarisation measurements towards star forming filaments have also revealed the presence of large-scale and rather well-ordered magnetic fields. In many of the observations the magnetic field appears to be roughly perpendicular to the filament \citep[e.g.][]{Chapman11,Sugitani11,Planck14}. However, \citet{Li13} report a bimodal distribution of magnetic field directions (parallel and perpendicular orientation) with respect to the major axis of the filamentary structure \citep[see also][]{Pillai14,Planck14b}. Moreover, by combining numerical simulations of magnetised, molecular clouds and synthetic polarisation maps, \citet{Soler13} show that the relative orientation of the magnetic field also depends on the initial magnetisation of the cloud in which the filaments are forming. The knowledge about the relative orientation would therefore allow to draw conclusions on the importance of magnetic fields during the formation of filaments: super-Alfv\'enic turbulence would cause strong compression, which results in magnetic fields parallel to the elongated structure \citep[e.g.][]{Padoan01}. In contrast, for a magnetic field guided gravitational contraction the field is preferentially perpendicular to the elongated structure \citep[e.g.][]{Nakamura08}. Furthermore, magnetic fields will also influence the subsequent evolution of filaments, a point we are particularly interested in in this work.

The stability and fragmentation properties of filaments with respect to various aspects like e.g. turbulent motions, external pressure confinement, or accretion onto the filament have been studied analytically in a number of papers \citep[e.g.][]{Ostriker64,Inutsuka92,Fischera12,Pon11,Pon12,Toala12,Heitsch13}. It has been found that magnetic fields generally have a positive effect on the stability of the filaments \citep{Nagasawa87,Fiege00,Heitsch13,Tomisaka14}. However, these studies have to make several simplifications like e.g. considering an infinitely long filament or treating turbulence as an additional pressure term.

There are also a number of numerical studies on the evolution and stability of filamentary structures. One of the first 2D numerical simulations were performed by \citet{Bastien83}, which demonstrate some typical fragmentation properties of filaments. Subsequently, more realistic 3D simulations showed the sensitivity of the fragmentation properties of filaments to the initial conditions like the mass per unit length, or the presence of global rotation or magnetic fields \citep[e.g.][]{Bastien91,Bonnell92,Tomisaka95}. More recently, \citet{Burkert04} report the formation of prominent clumps forming at the edge of elongated, collapsing structures \citep[see also][]{Clarke15}. On molecular cloud scales, the formation of star forming filaments has been studied by means of turbulent box simulations and colliding flows \citep[e.g.][]{MacLow04, Hennebelle08,Federrath10b,Gomez14,Moeckel14,Smith14,Kirk15}.

The present study systematically explores the impact of different magnetic field morphologies and turbulent motions on the evolution of star forming filaments by means of 3D MHD simulations. We find that for supercritical filaments a magnetic field perpendicular to the major axis has almost no effect on its stability, which results in rather narrow filaments. Only magnetic fields parallel to the major axis can stabilize the filament against radial collapse and reduce fragmentation. We will also show that turbulence strongly affects the fragmentation properties of filaments.

The paper is structured as follows: in Section~\ref{sec:IC}, we present an overview of the numerical methods and the initial conditions used in the simulations. Next, we present the results focussing on the impact of the different initial conditions, before we investigate the fragmentation, the radial density profiles, and general star formation properties of the filaments. In Section~\ref{sec:discussion} we discuss our findings and compare to recent observations and other numerical work, before we summarize in Section~\ref{sec:conclusions}.

\section{Numerical methods and initial conditions}
\label{sec:IC}

\subsection{Numerical methods}

We present 3D, magnetohydrodynamical (MHD) simulations of magnetized filaments using the astrophysical code FLASH 4.2.2~\citep{Fryxell00,Dubey08}. We solve the equations of ideal MHD on an adaptive mesh including self-gravity. The Poisson equation for gravity is solved using a multipole method based on a Barnes-Hut tree\footnote{implemented by R. W\"unsch, Academy of Sciences of the Czech Republic}. The employed MHD-solver preserves positive states and is well suited for highly supersonic, astrophysical problems~\citep{Bouchut07,Waagan09,Bouchut10,Waagan11}. The maximum spatial resolution, i.e. the size of the smallest grid cell is 40.3 AU. We apply a refinement criterion that guarantees that the Jeans length
\begin{equation}
 \lambda_\rmn{J} = \sqrt{\frac{\pi c_\rmn{s}^2}{G \rho}}
 \label{eq:jeans}
\end{equation}
is resolved with at least 16 grid cells. Here $c_s$ denotes the sound speed and $G$ the gravitational constant. The gas is isothermal at a temperature of 15 K, which agrees well with temperatures observed in a number of filaments \citep[e.g.][]{Andre10,Arzoumanian11,Li14}. The mean molecular weight is set to $\mu_\rmn{mol} = 2.3$ typical for molecular gas. In order to guarantee that the Jeans length is resolved everywhere, we create sink particles if the density exceeds a value of
\begin{equation}
 \rho_{\rmn{sink}} = 10^{-16} \, \rmn{g\,cm^{-3}}
 \label{eq:crit}
\end{equation}
\citep[for details on additional formation criteria see][]{Federrath10}. All gas in excess of that density, which is within a radius of 108 AU from the sink particle, is accreted.

\subsection{Initial conditions}

The properties of our filaments are motivated by number of recent observations \citep[e.g.][but see also \citealt{Andre13} for a recent review]{Andre10,Arzoumanian11,Peretto12,Busquet13,Palmeirim13,Schisano14}. We set the initial filament width to 0.1 pc and apply a Plummer-like density profile along the radial direction:
\begin{equation}
 \rho(r) = \frac{\rho_\rmn{c}}{\left[ 1 + (R/R_\rmn{flat})^2 \right]^{p/2}}
 \label{eq:fil}
\end{equation}
Here $R$ is the cylindrical radius, $\rho_\rmn{c}$ the central density, $R_\rmn{flat}$ the characteristic radius of the flat inner part of the density profile, and $p$ the characteristic exponent of the profile. We note that the Plummer profile is applied for the entire radial range, in particular no cut-off density is applied. \citet{Arzoumanian11} find typical values for $p$ between 1.3 and 2.4 with a mean of 1.6 and values for $R_\rmn{flat}$ between 0.01 pc and 0.08 pc with a mean of 0.03 pc, which gives a mean filament width of $3 \times$ $R_\rmn{flat} \sim 0.1$ pc. In this work we choose $p$ = 2 and $R_\rmn{flat} = 0.033$ pc. The length of the filaments is set to 1.6 pc. In order to avoid a pressure jump at the end of the filaments, the density decreases exponentially at each end of the filament.

In Table~\ref{tab:models} we list all simulations performed and their corresponding parameters. 
\begin{table*}
  \caption{Performed simulations of filaments with a length of 1.6 pc and a width 0.1 pc. We list the run name, the central density $\rho_\rmn{c}$, the mass per unit length $(M/L)_\rmn{fil}$, the magnetic field configuration (with respect to the major axis of the filament) and strength, the Mach number, and the turbulent integral scale $\lambda_\rmn{max}$.}
 \label{tab:models}
 \begin{tabular}{@{}lcccccc}
  \hline
  Run & $\rho_{\rmn{c}}$ [g cm$^{-3}$] & $(M/L)_\rmn{fil}$ [M$_{\sun}$/pc] & B-field & $B$ [$\mu$G] & Mach number & $\lambda_\rmn{max}$ [pc] \\
  \hline
  F1\_NoMag\_L2 & 1 $\times$ 10$^{-19}$ & 25 & no field & 0 & 1 & 0.1 \\
  F1\_para\_L2 & 1 $\times$ 10$^{-19}$& 25 &  parallel & 40 & 1 & 0.1 \\
  F1\_perp\_L2 & 1 $\times$ 10$^{-19}$& 25 &  perpendicular & 40 & 1 & 0.1 \\
  F1\_perp\_L2\_Bweak & 1 $\times$ 10$^{-19}$& 25 & perpendicular & 10 & 1 & 0.1\\
  \hline
  F2\_NoMag\_L2 & 1.4 $\times$ 10$^{-19}$& 35 &  no field & 0 & 1 & 0.1 \\
  F2\_perp\_2 & 1.4 $\times$ 10$^{-19}$& 35 &  perpendicular & 40 & 1 & 0.1 \\
  \hline
  F3\_NoMag\_L2 & 3 $\times$ 10$^{-19}$& 75 &  no field & 0 & 1 & 0.1 \\
  F3\_para\_L2 & 3 $\times$ 10$^{-19}$& 75 &  parallel & 40 & 1 & 0.1 \\
  F3\_perp\_L2 & 3 $\times$ 10$^{-19}$& 75 &  perpendicular & 40 & 1 & 0.1 \\
  \hline
  F3\_NoMag\_L1 & 3 $\times$ 10$^{-19}$& 75 &  no field & 0 & 1 & 0.0375 \\
  F3\_NoMag\_L3 & 3 $\times$ 10$^{-19}$& 75 &  no field & 0 & 1 & 0.3 \\
  F3\_NoMag\_L2\_M2.5 & 3 $\times$ 10$^{-19}$& 75 &  no field & 0 & 2.5 & 0.1 \\
  F3\_NoMag\_L2\_condensed$^a$ & 3 $\times$ 10$^{-19}$ & 75$^a$ & no field & 0 & 1 & 0.1 \\
  \hline
  F3\_para\_L1 & 3 $\times$ 10$^{-19}$& 75 &  parallel & 40 & 1 & 0.0375\\
  F3\_para\_L3 & 3 $\times$ 10$^{-19}$& 75 &  parallel & 40 & 1 & 0.3 \\
  F3\_para\_L2\_Brad$^b$ & 3 $\times$ 10$^{-19}$& 75 &  parallel & 40$^b$ & 1 & 0.1 \\
  \hline
 \end{tabular}
 \\
$^a$ central density enhanced by a factor of 3, see equation~\ref{eq:enhance} \\
$^b$ radial dependence of the magnetic field, see equation~\ref{eq:dropoff} \\
\end{table*}
An important parameter for the evolution of a filament is its mass per unit length. As shown by \citet{Ostriker64}, there is a critical mass per unit length
\begin{equation}
 (M/L)_\rmn{crit} = \frac{2 c_\rmn{s}^2}{G} \, ,
\end{equation}
above which the filament starts to collapse along its radial direction and gets prone to gravitationally induced fragmentation ($(M/L)_\rmn{crit} \sim$ 25 M$_{\sun}$/pc for gas with 15 K). In this work we investigate the evolution of filaments, which have $(M/L)_\rmn{fil}$ = 1, 1.4, and 3 $\times$ $(M/L)_\rmn{crit}$, i.e. $(M/L)_\rmn{fil}$ = 25, 35, and 75 M$_{\sun}$/pc (runs labelled with ``F1'', ``F2'', and ``F3'', respectively, see Table~\ref{tab:models}). Given the density profile in equation~\ref{eq:fil}, this results in a central density of $\rho_\rmn{c} = 1\times10^{-19}$, $1.4\times10^{-19}$, and $3\times10^{-19}$ g cm$^{-3}$, respectively.
According to \citet{Arzoumanian11}, this can be converted into central column densities
\begin{equation}
 \Sigma_\rmn{c} = A_\rmn{p} \rho_\rmn{c} R_\rmn{flat}
\end{equation}
with $A_\rmn{p} = \frac{1}{\rmn{cos \, i}} \int_{-\infty}^{\infty} \frac{\rmn{d}u}{(1+u^2)^{p/2}}$. Assuming an inclination angle $i$ = 0, this  results in $\Sigma_\rmn{c}$ between $8.4 \times 10^{21}$ and $25.2 \times 10^{21}$ cm$^{-2}$, comparable to the authors findings.

Observations show that filaments are most likely magnetized with a direction of the magnetic field which is either perpendicular to the major axis\footnote{Throughout the paper we will use \textit{major} axis for the long (symmetry) axis of the filament and \textit{radial} for the direction perpendicular to it.} of the filament \citep[e.g.][]{Chapman11,Sugitani11,Planck14} or parallel to it \citep{Li13,Pillai14,Planck14b}. The preferred direction of the magnetic field in the densest parts, however, is still not entirely certain: Dust polarisation maps usually reveal a strong decrease of the polarisation towards the densest regions, which leads to some uncertainty about the field direction within the filament \citep[e.g.][]{Hildebrand99,Hull14}. Furthermore, as shown by \citet{Soler13}, also the initial magnetisation of the surrounding molecular cloud can have a significant impact on the relative orientation of the magnetic field. Therefore, we will test both extreme cases here, i.e. perpendicular and parallel magnetic fields (run names contain ``perp'' or ``para''; runs without a magnetic field are termed ``NoMag''). We take the magnetic field strength to be 40 $\mu$G, which results in a magnetic pressure in the centre of the filament comparable to the thermal pressure, i.e.
\begin{equation}
 c_\rmn{s} \simeq v_\rmn{A} \, ,
 \label{eq:equipartition}
\end{equation}
where $v_\rmn{A}$ is the Alfv\'en velocity
\begin{equation}
 v_\rmn{A} = \frac{B}{\sqrt{4 \pi \rho}} \, .
 \label{eq:Alfven}
\end{equation}
In the case of a perpendicular magnetic field this corresponds to a normalized mass-to-flux ratio between 1.6 and 4.8 for the different simulations. Despite the fact that the choice of 40 $\mu$G is somewhat arbitrary, it compares reasonably well with recent estimates of magnetic field strengths within filaments \citep[e.g.][]{Alves08,Chapman11,Sugitani11}. Moreover, as shown by \citet[see their figures 6 and 7]{Crutcher12} at a volume particle density of $n \sim 10^4 - 10^5$ cm$^{-3}$ or a column density around $10^{22}$ cm$^{-2}$ as present in our filaments, the corresponding magnetic field strength in the galactic interstellar medium is of the order of a few 10 $\mu$G, which agrees well with our choice. The magnetic field is initially constant throughout the simulation domain. Only for run F1\_perp\_L2\_Brad, the magnetic field decreases along the radial direction according to \citep{Crutcher12}
\begin{equation}
 B(R) = B_0 \, \left(\frac{\rho(R)}{\rho_\rmn{c}}\right)^{0.5} \, ,
 \label{eq:dropoff}
\end{equation}
with the cylindrical radius $R$ and $B_0 = 40$ $\mu$G.

Observations of star-forming filaments reveal non-thermal line widths, which indicate an additional turbulent velocity field with typical Mach numbers between 1 and 3 \citep{Arzoumanian13,Hacar13,Furuya14,Henshaw14,Jimenez14,Li14}. For this reason, we apply a transonic or supersonic turbulent velocity field. We use a power-law spectrum with a slope of -11/3, corresponding to a Kolmogorov type spectrum in 3D. For each of the simulations, we choose a different random seed for the turbulence field. The typical length of the largest velocity fluctuations, i.e. the integral scale $\lambda_\rmn{max}$ of the turbulence field, is 0.1 pc, thus identical to the filament width. We change the turbulent Mach number and $\lambda_\rmn{max}$, which is indicated by the run name (see Table~\ref{tab:models}), where ``L1'', ``L2'', and ``L3'' are used for $\lambda_\rmn{max} = 0.0375, 0.1$, and 0.3 pc, respectively. In all simulations the smallest scale for velocity fluctuations is 80 AU, i.e. 2 grid cells, the Mach number (with respect to the sound speed) in all runs but one is 1. This is at the low side but still in agreement with the observational results mentioned above. We also performed an additional run with a Mach number of 2.5 (run F3\_NoMag\_L2\_M2.5). We note that the Alfv\'en Mach number is comparable to the (sonic) Mach number (see Equation~\ref{eq:equipartition}).

For all runs but one (run F3\_NoMag\_L2\_condensed) the density is constant \textit{along} the major axis of the filament. For run F3\_NoMag\_L2\_condensed we enhance the density towards the centre of the filament by a factor of three:
\begin{equation}
 \rho(r,x) =  \rho_\rmn{c} \times \left( \frac{1}{1+(R^2/R_\rmn{flat})^2)} + \frac{2}{1 + (R^2 + x^2)/R_\rmn{flat}^2)} \right) \, ,
 \label{eq:enhance}
\end{equation}
where $R$ is the cylindrical radius, $x$ the distance along the major axis, $\rho_\rmn{c} = 3 \times 10^{-19}$ g cm$^{-3}$, and the exponent $p$ set to 2. In this run the magnetic field is zero, the turbulent Mach number is 1, and $\lambda_\rmn{max} = 0.1$ pc.

In order to minimize boundary effects from the simulation domain, we place the filament in the centre of a domain which extents over 0.8 pc in the $y$- and $z$-direction and 2.4 pc in the $x$-direction. Thus, the effective maximum resolution is $4096^2 \times 12288$ cells. We use isolated boundaries for the gravitational potential and open (Dirichlet) boundary conditions for the gas.

\section{Results}

In this section we present the results of the simulations. First (Section~\ref{sec:timeevol}), we focus on the time evolution of three fiducial runs F1\_para\_L2, F3\_para\_L2, and F3\_perp\_L2. Next, we analyse the dependence of the fragmentation mode on global filament properties (Section~\ref{sec:fragmode}) before we consider the evolution of the radial density profiles (Section~\ref{sec:profiles}). In Section~\ref{sec:frag}, we investigate the distances between the fragments in the filaments, the global star formation rates are discussed in Section~\ref{sec:SF}.

We evolve each simulation for $t_\rmn{evol} =$ 100 kyr -- 250 kyr \textit{after} the first protostar has formed  at $t_\rmn{form}$. We follow each simulation over a long enough time to be able to draw reliable conclusions about properties like fragmentation and star formation rates. For this reason, $t_\rmn{evol}$ is different for each simulation. Hence, also the final simulation times $t_\rmn{end} = t_\rmn{form} + t_\rmn{evol}$ change. For run F1\_perp\_L2, there is no star formation at all within 2 Myr and we stop the run at this point.

The formation time of the first protostar varies from 95 kyr up to 1.4 Myr depending on the mass, the magnetic field configuration, and turbulence strength of the simulated filaments. Using the central gas density $\rho_\rmn{c}$, the resulting free-fall times
\begin{equation}
 t_\rmn{ff} = \sqrt{\frac{3 \pi}{32 G \rho_\rmn{c}}}
\end{equation}
are 210, 178, and 122 kyr for filaments with ($M/L)_\rmn{fil}$ = 25, 35, and 75 M$_{\sun}$/pc, respectively (86 kyr for run F3\_NoMag\_L2\_condensed). As can be seen from Table~\ref{tab:results}, where all time scales are listed, $t_\rmn{form}$ is larger than $t_\rmn{ff}$ by up to a factor of about 7 (run F1\_perp\_L2\_Bweak). This is due to the fact that in particular for the filaments with ($M/L)_\rmn{fil}$ = 25 M$_{\sun}$/pc the thermal pressure substantially counteracts gravity and free-fall is not a valid assumption any more. Also for the filaments with ($M/L)_\rmn{fil}$ = 75 M$_{\sun}$/pc, $t_\rmn{form}$ is larger than $t_\rmn{ff}$ by up to a factor of $\sim$ 3, which we attribute to the combined effect of thermal and magnetic pressure. We note that using the mean density of the filaments within a cylindrical radius of 0.2 pc, which is about a factor of 10 lower than $\rho_\rmn{c}$, the corresponding $t_\rmn{ff}$ is larger by a factor of $\sim$ 3, which generally results in a better agreement between $t_\rmn{form}$ and $t_\rmn{ff}$.

Due to the open boundaries, gas is allowed to flow into the simulation domain. The total mass inflow rates of the different simulations range from $10^{-6}$ to $10^{-5}$ M$_{\sun}$ yr$^{-1}$. Since these rates are about one order of magnitude smaller than the accretion rates onto the sink particles (see Section~\ref{sec:SF}), we can neglect the impact of this additional mass increase (a few per cent) in the box.

\begin{table*}
\centering
  \caption{Free-fall time $t_\rmn{ff}$ calculated using the central density $\rho_\rmn{c}$, formation time $t_\rmn{form}$ of the first protostar, simulation time $t_\rmn{evol}$ elapsed after $t_\rmn{form}$, total simulated time $t_\rmn{end}$, total time-averaged accretion rate, fragmentation mode, and number of fragments at $t_\rmn{end}$.}
 \label{tab:results}
 \begin{tabular}{@{}lcccccccc}
  \hline
  Run & $t_\rmn{ff}$ [kyr] & $t_\rmn{form}$ [kyr] & $t_\rmn{evol}$ [kyr] & $t_\rmn{end}$ [kyr] & $\dot{M}_\rmn{star}$ [10$^{-4}$ M$_{\sun}$ yr$^{-1}$] & fragmentation mode& \# fragments\\
  \hline
  F1\_NoMag\_L2 & 210 & 1309 & 250 & 1559 & 1.11 & edge-on & 3 \\
  F1\_para\_L2 & 210 & 1122 & 250 & 1373 & 0.391 & edge-on & 2 \\
  F1\_perp\_L2 & 210 & --- & --- & 2000 & 0 & --- & 0 \\
  F1\_perp\_L2\_Bweak & 210 & 1398 & 250 & 1648 & 0.602 & edge-on & 3 \\
  \hline
  F2\_NoMag\_L2 & 178 & 692 & 200 & 893 & 1.70 & mixed & 18 \\
  F2\_perp\_2 & 178 & 1239 & 150 & 1389 & 0.43 & edge-on & 2 \\
  \hline
  F3\_NoMag\_L2 & 122 & 239 & 150 & 389 & 5.01 & mixed & 24 \\
  F3\_para\_L2 & 122 & 282 & 150 & 432 & 1.30 & mixed & 7 \\
  F3\_perp\_L2 & 122 & 264 & 150 & 414 & 3.60 & uniform & 120 \\
  \hline
  F3\_NoMag\_L1 & 122 & 240 & 110 & 350 & 7.78 & mixed & 135 \\
  F3\_NoMag\_L3 & 122 & 163 & 200 & 363 & 2.78 & mixed & 17 \\
  F3\_NoMag\_L2\_M2.5 & 122 & 293 & 100 & 393 & 6.13 & mixed & 28 \\
  F3\_NoMag\_L2\_condensed & 86 & 96 & 150 & 246 & 2.34 & centralised & 14 \\
  \hline
  F3\_para\_L1 & 122 & 290 & 150 & 440 & 1.12 & edge-on & 2 \\
  F3\_para\_L3 & 122 & 179 & 200 & 379 & 1.52 & mixed & 6 \\
  F3\_para\_L2\_Brad & 122 & 284 & 100 & 384 & 2.57 & mixed & 9 \\
  \hline
 \end{tabular}
\end{table*}

\subsection{Time evolution}
\label{sec:timeevol}

\begin{figure*}
 \includegraphics[width=\linewidth]{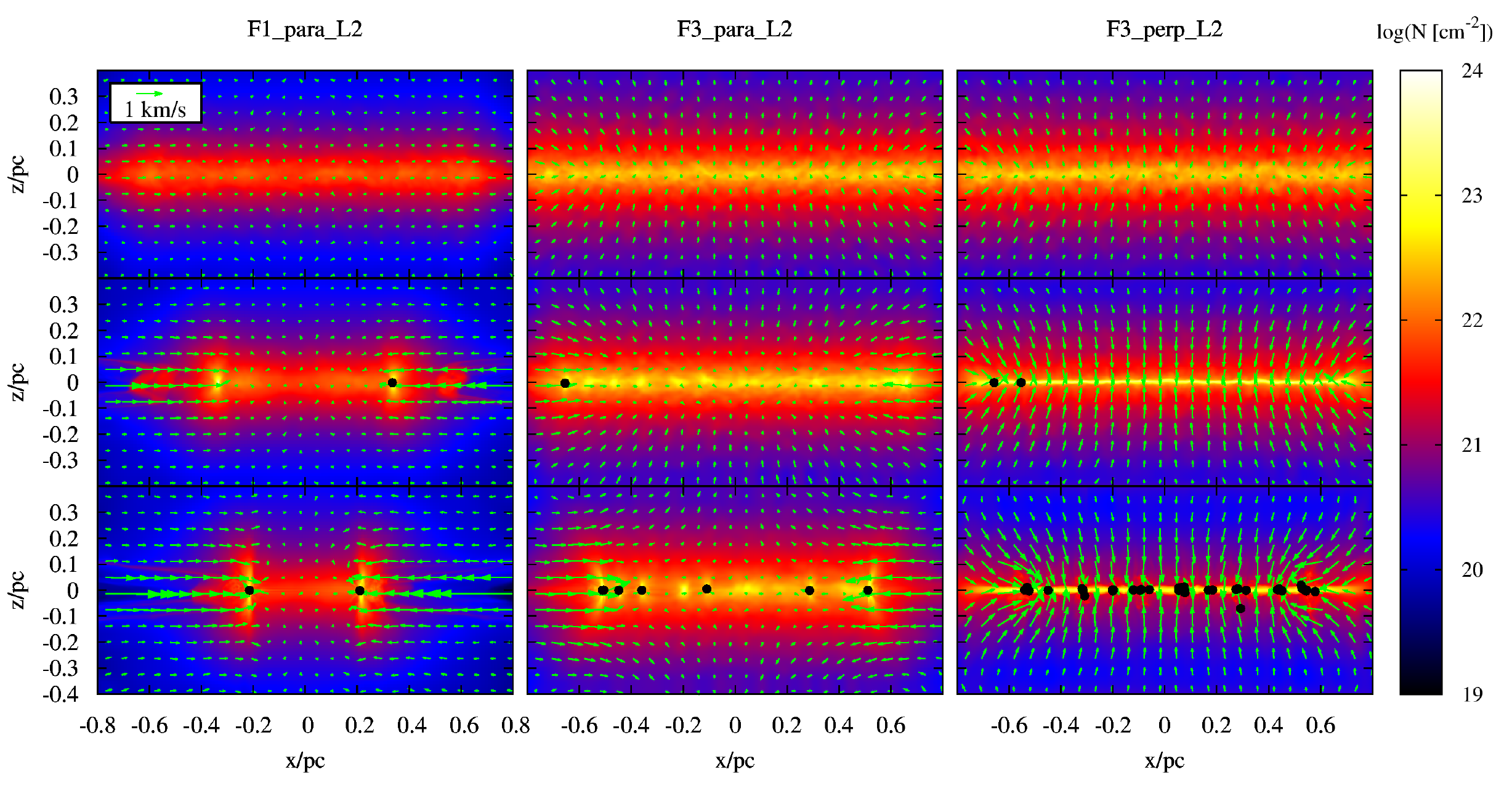}
 \caption{Gas column density and line-of-sight averaged velocity field (green arrows) at $t$ = $t_\rmn{form}/2$, $t_\rmn{form}$, and $t_\rmn{end}$ (from top to bottom) for the runs F1\_para\_L2, F3\_para\_L2, and F3\_perp\_L2 (from left to right). Black dots represent sink particles. The left and middle column demonstrates the impact of varying filament masses, whereas the middle and right column demonstrate the impact of different magnetic field morphologies. Note that the snapshots in each row are not taken at identical physical times but at comparable evolutionary stages (see Table~\ref{tab:results} for the actual physical times).}
 \label{fig:timeevol}
\end{figure*}

First, we consider the time evolution of the runs F1\_para\_L2, F3\_para\_L2, and F3\_perp\_L2, which demonstrate some basic differences that occur in the evolution of filaments with different magnetic field morphologies and values of ($M/L)_\rmn{fil}$. In Fig.~\ref{fig:timeevol} we show the column density of the three runs at $t = t_\rmn{form}/2$, $t_\rmn{form}$, and $t_\rmn{end}$.

For run F1\_para\_L2 (left column of Fig.~\ref{fig:timeevol}), the filament starts to contract along its major axis towards its geometrical centre whereas is remains stable along its radial direction (see also Section~\ref{sec:profiles} for more details). At the outer edges of the filament overdense condensations form, which in turn collapse to form sink particles at $t_\rmn{form} = 1112$ kyr. Over time, the sinks move towards the centre of the filament. Such a behaviour was previously found in numerical simulations by \citet{Burkert04} and was explained (semi-) analytically by a number of authors \citep{Pon11,Pon12,Clarke15}. Interestingly, ahead of the condensation, i.e. at the side towards the centre, the gas is accelerated {\it outwards} in the direction of the clump, in agreement with the findings of \citet{Clarke15}. Considering the inner parts of the filament, no further fragmentation occurs. Throughout the paper we will call this fragmentation behaviour, which is dominated by two fragments at each end of the filament, an \textit{edge-on} collapse mode.

Next, we consider the time evolution of run F3\_para\_L2, which has ($M/L)_\rmn{fil} = 75 M_{\sun}$/pc $\simeq 3 (M/L)_\rmn{crit}$ (middle column of Fig.~\ref{fig:timeevol}). Also in this run, the column density profile does not appear to change significantly over time. However, due to the higher mass, the first sink particle forms significantly earlier ($t_\rmn{form}$ = 282 kyr) than in run F1\_para\_L2, although, also in this run, the two first particles form at each end of the filament. However, more sink particles subsequently form along the major axis of the filament, which is why we call this collapse mode \textit{mixed}.

In the right column of Fig.~\ref{fig:timeevol} we show the time evolution of run F3\_perp\_L2, which has a magnetic field perpendicular to the major axis of the filament. Heavy fragmentation occurs along the filament with a significantly larger number of fragments compared to run F3\_para\_L2 (see Table~\ref{tab:results}). This is due to the lack of an additional magnetic pressure which stabilizes the filament against radial contraction in the first two simulations considered. Consequently, the filament in run F3\_perp\_L2 is much thinner and has a higher central column density. In run F3\_perp\_L2 the collapse does not proceed in an edge-on fashion. Fragmentation rather occurs randomly along the entire filament, with the first sink particles forming closer to the centre. We call this fragmentation mode \textit{uniform}.

The different fragmentation modes also impact the distribution of fragment masses. For run F1\_para\_L2 the two sink particles have quite similar masses of 5.3 and 4.5 M$_{\sun}$. For run F3\_para\_L2 the two most massive particles are the outermost particles with masses of 7.4 and 5.9 M$_{\sun}$, whereas the remaining five particles have smaller masses between 0.9 and 1.4 M$_{\sun}$. For run F3\_perp\_L2, the situation is quite different. Here only one massive particle with 4.6 M$_{\sun}$ exists. Furthermore, there are 10 more sinks with masses between 1 and 3.3 M$_{\sun}$, 102 with masses between 0.1 and 1 M$_{\sun}$, and 7 light particles with masses below 0.1 M$_{\sun}$. We emphasize that in general runs with the same fragmentation mode (see Table~\ref{tab:results}) also reveal a similar distribution of protostellar masses. The actual distribution, however, might also depend on the spatial resolution of the simulation. In particular, we would expect a larger number of lighter particles with increasing resolution. However, we tentatively expect the general trend described above to hold even for a higher spatial resolution.

Overall, we demonstrate the impact of different initial properties like the filament mass and magnetic field morphology on the fragmentation mode, the number of fragments, the filament width, and the distribution of fragment masses. We find that even small changes in the initial configuration of the filament can cause characteristic differences. In the following sections we discuss these differences in more detail.

\subsection{Fragmentation mode}
\label{sec:fragmode}

\begin{figure*}
 \includegraphics[width=\linewidth]{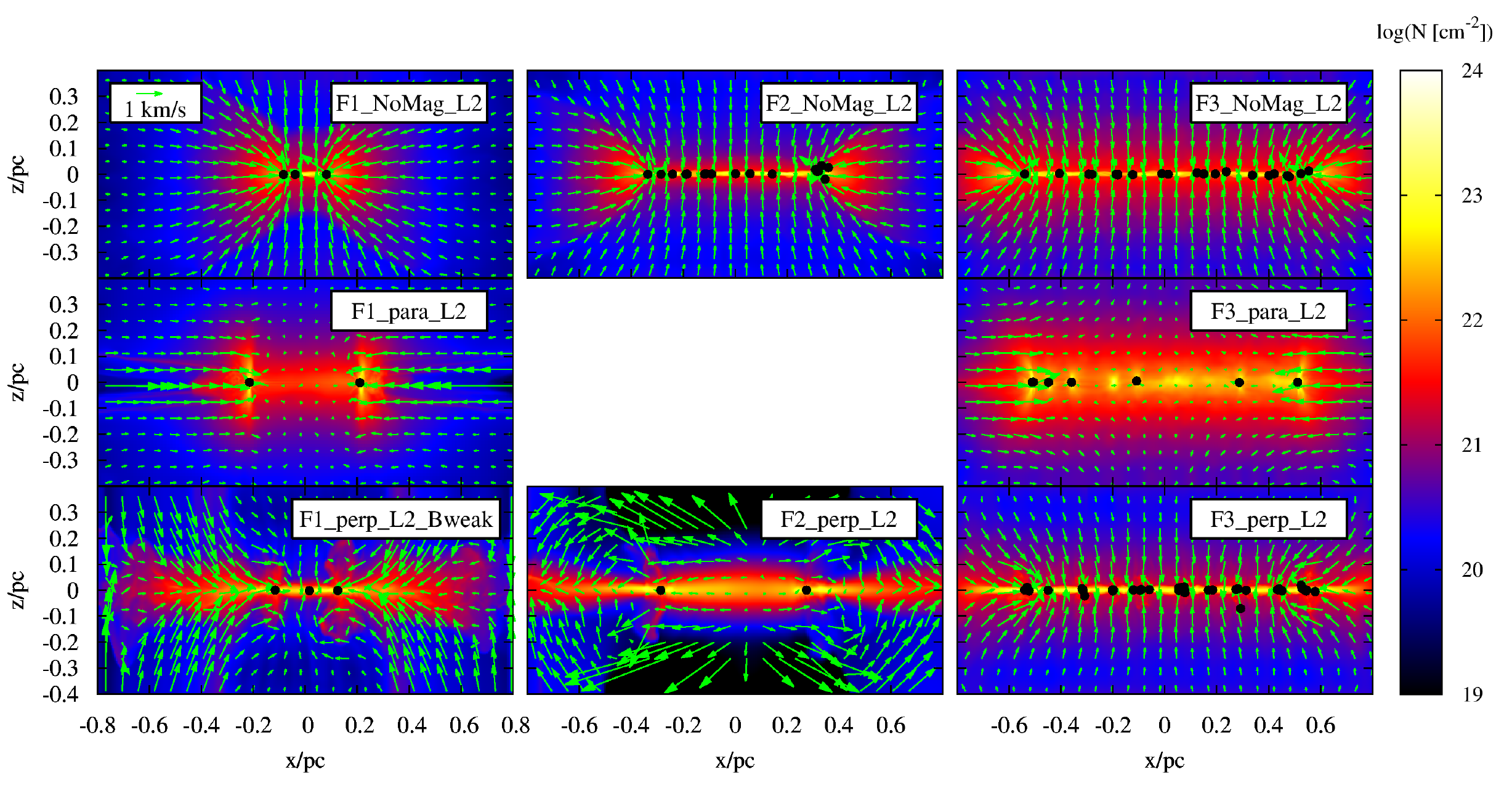}
 \caption{Gas column density and line-of-sight averaged velocity field (green arrows) at $t$ = $t_\rmn{end}$ for the runs with  $\lambda_\rmn{max} = 0.1$ pc and a turbulent Mach number of 1. Black dots represent sink particles. The figure demonstrates the impact of the mass per length of the filament and the orientation of the magnetic field on the fragmentation properties.}
 \label{fig:frag}
\end{figure*}

In this section, we discuss the effect of the initial conditions on the fragmentation mode for all simulations with a fixed turbulent integral scale $\lambda_\rmn{max} = 0.1$ pc and a turbulent Mach number of 1 (see Table~\ref{tab:models}). Here, we only consider the final state ($t_\rmn{end}$) since usually the time evolution of the individual runs is qualitatively similar to one of the three fiducial runs discussed in Section~\ref{sec:timeevol}. We will show that filaments with $(M/L)_\rmn{fil} = (M/L)_\rmn{crit}$ preferentially collapse in an edge-on fashion. For more massive filaments this is not the case, and only longitudinal magnetic fields help to stabilize the filament against heavy fragmentation. An overview showing all simulation discussed below at $t$ = $t_\rmn{end}$ is given in Fig.~\ref{fig:frag}.

\subsubsection{Filaments with $(M/L)_\rmn{fil} = (M/L)_\rmn{crit}$}

Similar to run F1\_para\_L2, in simulation F1\_NoMag\_L2 the formation of the sink particles initially occurs at the edges of the filament although a third sink particle forms about 229 kyr after the first sink particle in the inner region, approximately 6100 AU offset from the centre. The overall similar behaviour between the both runs concerning the fragmentation properties is not surprising since both have $(M/L)_\rmn{fil} = (M/L)_\rmn{crit}$ and are thus marginally stable against fragmentation even without the presence of a magnetic field. Interestingly, the first sink particle in run F1\_para\_L2 is formed somewhat earlier than in run F1\_NoMag\_L2 ( about 15\% of $t_\rmn{form}$, see Table~\ref{tab:results}), which could be due to the magnetic field, which channels the gas along the longitudinal direction resulting in a somewhat faster accumulation of mass at the edges of the filament.

In run F1\_perp\_L2, where the magnetic field is oriented perpendicular to the filament, the formation of sink particles is completely suppressed up to $t_\rmn{end}$ = 2 Myr at which we stop the simulation. Here, the additional magnetic pressure stabilizes the marginally critical filament against gravitational collapse, which is in good agreement with a mass-to-flux ratio of 1.6, i.e. a filament which is only marginally magnetically supercritical. However, in order to explore the stabilizing effect of the magnetic field in more detail, we repeat this simulation with a four times weaker magnetic field of 10 $\mu$G (run F1\_perp\_L2\_Bweak). As listed in Table~\ref{tab:results}, star formation now occurs at $t_\rmn{form} = 1398$ kyr, i.e. approximately at the same time as in the runs F1\_NoMag\_L2 and F1\_para\_L2. Moreover, also for this run collapse occurs in an edge-on fashion with a third fragment forming 244 kyr after the first sink particle about 3200 AU away from its geometrical centre.

\subsubsection{Filaments with $(M/L)_\rmn{fil} = 1.4 \times (M/L)_\rmn{crit}$}

Next, we analyse the fragmentation mode of the runs F2\_NoMag\_L2 and F2\_perp\_2 which have a mass per length of 35 M$_{\sun}$/pc $\simeq 1.4 (M/L)_\rmn{crit}$. For run F2\_NoMag\_L2 the fragmentation mode is \textit{mixed} where the fragments form first at the edges and subsequently along the entire filament. For run F2\_perp\_2, fragmentation is still dominated by an edge-on collapse and no further sink particles form along the filament up to $t_\rmn{end}$. It appears that even for marginally supercritical filaments ($(M/L)_\rmn{fil} \leq 1.4 \times (M/L)_\rmn{crit}$) a magnetic field perpendicular to its major axis can stabilize it against heavy fragmentation or even completely suppress star formation as in run F1\_perp\_L2. We note that in the outer regions of run F1\_perp\_L2 the magnetic pressure has started to push material outwards causing the peculiar velocity field structure.

\subsubsection{Filaments with $(M/L)_\rmn{fil} = 3 \times (M/L)_\rmn{crit}$}

In run F3\_NoMag\_L2, fragmentation occurs in a mixed mode, where the first sink particles form at either end of the filament. The number of fragments is significantly higher than in run F3\_para\_L2 (7 vs. 24 fragments) which we attribute to the stabilizing effect of the magnetic field in the latter case. Interestingly, compared to run F3\_perp\_L2, fragmentation seems to be be somewhat reduced in run F3\_NoMag\_L2. Comparing the kinetic energy content of F3\_perp\_L2 and F3\_NoMag\_L2 (not shown here), we find that in the former case turbulence decays somewhat more quickly in the beginning. Hence, we argue that the turbulent pressure counteracting gravity is smaller, which in turn results in a higher number of fragments in run F3\_perp\_L2. 

To summarize, increasing the mass per length of a filament results in an enhanced fragmentation over the entire extent of the filament. A magnetic field parallel to the filament helps to stabilize it against radial collapse and subsequent fragmentation. For perpendicular magnetic fields, only for marginally supercritical filaments ($(M/L)_\rmn{fil} \leq 1.4 (M/L)_\rmn{crit}$) the field can contribute to the stabilization of the filament.

\subsection{Density profiles}
\label{sec:profiles}

Initially, the filaments have a radial density profile in agreement with recent observations (see Section~\ref{sec:IC}). In the following we analyse how the radial profiles evolve and whether their functional form is retained.

\begin{figure*}
 \includegraphics[width=\linewidth]{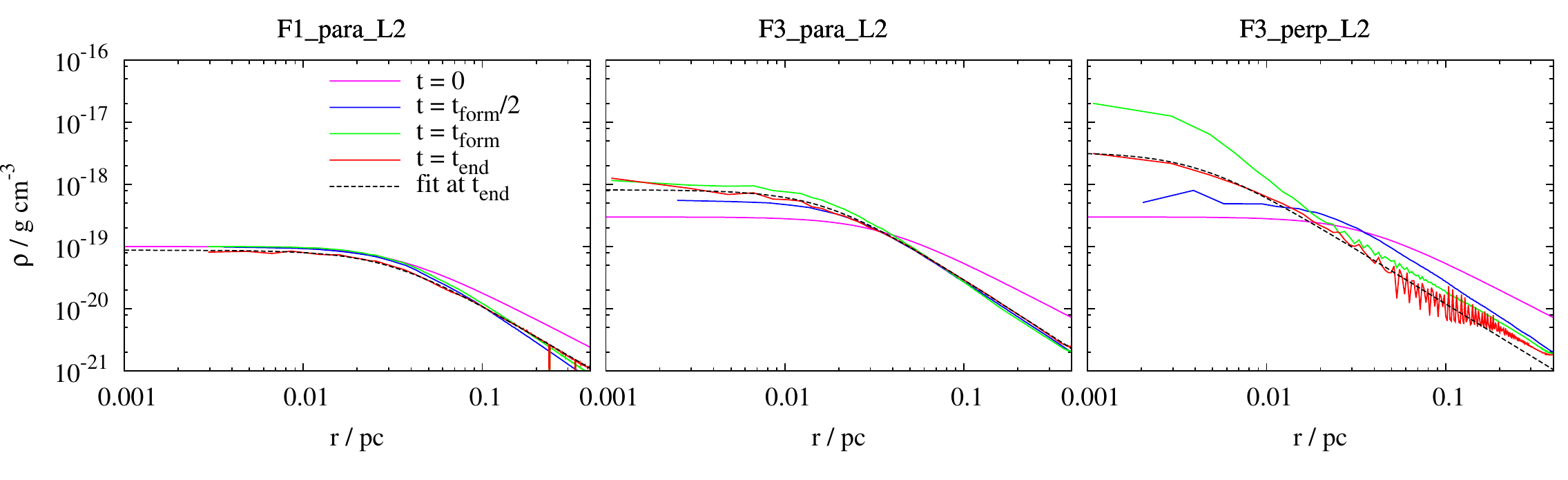}
 \caption{Averaged density profile along the radial direction of the filament for the runs F1\_para\_L2, F3\_para\_L2, and F3\_perp\_L2 (from left to right). The times are identical to those in Fig.~\ref{fig:timeevol} (see also text). The dashed black lines represent the best fit (equation~\ref{eq:fil}) at $t = t_\rmn{end}$ (red lines).}
 \label{fig:profile}
\end{figure*}

We display the radial density profiles for runs F1\_para\_L2, F3\_para\_L2, and F3\_perp\_L2 in Fig.~\ref{fig:profile} at four different times ($t$ = $0$, $t_\rmn{form}/2$, $t_\rmn{form}$, and $t_\rmn{end}$). In order to reduce local fluctuations, we average the density along the major axis, excluding the edges, which have already fragmented. For run F1\_para\_L2 we use the inner 0.8 pc (i.e. 0.4 pc in either direction from the centre) at $t = t_\rmn{form}/2$, 0.4 pc at $t = t_\rmn{form}$, and 0.2 pc at $t = t_\rmn{end}$; for the runs F3\_para\_L2 and F3\_perp\_L2 we take the inner 1.2 pc, 1.0 pc, and 0.8 pc at $t = t_\rmn{form}/2$, $t_\rmn{form}$, and $t_\rmn{end}$, respectively.

For run F1\_para\_L2 (left panel), the density profile remains similar to the initial profile. This is also valid for run F3\_para\_L2 (middle panel) despite a small increase in the central density. However, in run F3\_perp\_L2, the central density increases by about one order of magnitude. This indicates that here the thermal pressure of the gas is not sufficient to stabilize the filament against radial collapse, in agreement with Fig.~\ref{fig:timeevol}. We note that the central density at $t_\rmn{form}$ (green line) is somewhat higher than at $t_\rmn{end}$ since part of the gas is accreted onto the sink particles, whose mass we do not take into account when calculating the density profiles at later stages.

We fit equation~\ref{eq:fil} to the density profiles shown in Fig.~\ref{fig:profile}. The resulting values for $R_\rmn{flat}$, $p$, and $\rho_\rmn{c}$ are listed in Table~\ref{tab:fits}. In addition, we show the fit for $t = t_\rmn{end}$ (black dashed lines) in Fig.~\ref{fig:profile}.
\begin{table}
\centering
  \caption{Best fit parameter $R_\rmn{flat}$, $p$, and $\rho_\rmn{c}$ from equation~\ref{eq:fil} for the radial density profiles of selected runs.}
 \label{tab:fits}
 \begin{tabular}{@{}lccc}
  \hline
  Run & $R_\rmn{flat}$ [pc] & $p$ & $\rho_\rmn{c}$ [g cm$^{-3}$] \\
  \hline
  F1\_NoMag\_L2, $t$ = $t_\rmn{end}$ & 0.011 & 10 & 101 $\times$ 10$^{-19}$ \\
  \hline
  F1\_para\_L2, $t$ = $t_\rmn{form}/2$ & 0.036 &  2.1 & 1.02 $\times$ 10$^{-19}$  \\
  F1\_para\_L2, $t$ = $t_\rmn{form}$ & 0.037 &  2.0 & 1.05 $\times$ 10$^{-19}$  \\
  F1\_para\_L2, $t$ = $t_\rmn{end}$ & 0.030 &  1.7 & 0.87 $\times$ 10$^{-19}$  \\
  \hline
  F1\_perp\_L2, $t$ = $t_\rmn{end}$ & 0.028 & 2.1 & 4.4 $\times$ 10$^{-19}$ \\
  \hline
  F2\_NoMag\_L2, $t$ = $t_\rmn{end}$ & 0.0075  & 5.6 & 96 $\times$ 10$^{-19}$ \\
  F2\_perp\_2, $t$ = $t_\rmn{end}$ & 0.043  & 3.4 & 2.9 $\times$ 10$^{-19}$ \\
  \hline
  F3\_NoMag\_L2, $t$ = $t_\rmn{end}$ & 0.015 & 12 & 45 $\times$ 10$^{-19}$ \\ 
  \hline
  F3\_para\_L2, $t$ = $t_\rmn{form}/2$ & 0.021 & 1.9 & 6.0 $\times$ 10$^{-19}$ \\
  F3\_para\_L2, $t$ = $t_\rmn{form}$ & 0.016 & 2.0 & 11.2 $\times$ 10$^{-19}$ \\
  F3\_para\_L2, $t$ = $t_\rmn{end}$ & 0.017 & 1.9 & 8.2 $\times$ 10$^{-19}$ \\
  \hline
  F3\_perp\_L2, $t$ = $t_\rmn{form}/2$ & 0.022 & 2.0 & 6.0 $\times$ 10$^{-19}$ \\
  F3\_perp\_L2, $t$ = $t_\rmn{form}$ & 0.0014 & 1.8 & 385 $\times$ 10$^{-19}$ \\
  F3\_perp\_L2, $t$ = $t_\rmn{end}$ & 0.0041 & 1.8 & 33 $\times$ 10$^{-19}$ \\
  \hline
  F3\_NoMag\_L2\_M2.5, $t$ = $t_\rmn{end}$ & 0.0090 & 5.8 & 86 $\times$ 10$^{-19}$ \\
  \hline
 \end{tabular}
\end{table}
For the runs F1\_para\_L2 and F3\_para\_L2 (left and middle panel), the profiles evolve only little  and the fit parameters are comparable to the initial values and therefore agree reasonably well with those of \citet{Arzoumanian11}. For run F3\_perp\_L2 (right panel of Fig.~\ref{fig:profile}), only the fit at $t$ = $t_\rmn{form}/2$ is still similar to the initial one, whereas we obtain a much more narrow profile at later times.

\begin{figure}
 \includegraphics[width=\linewidth]{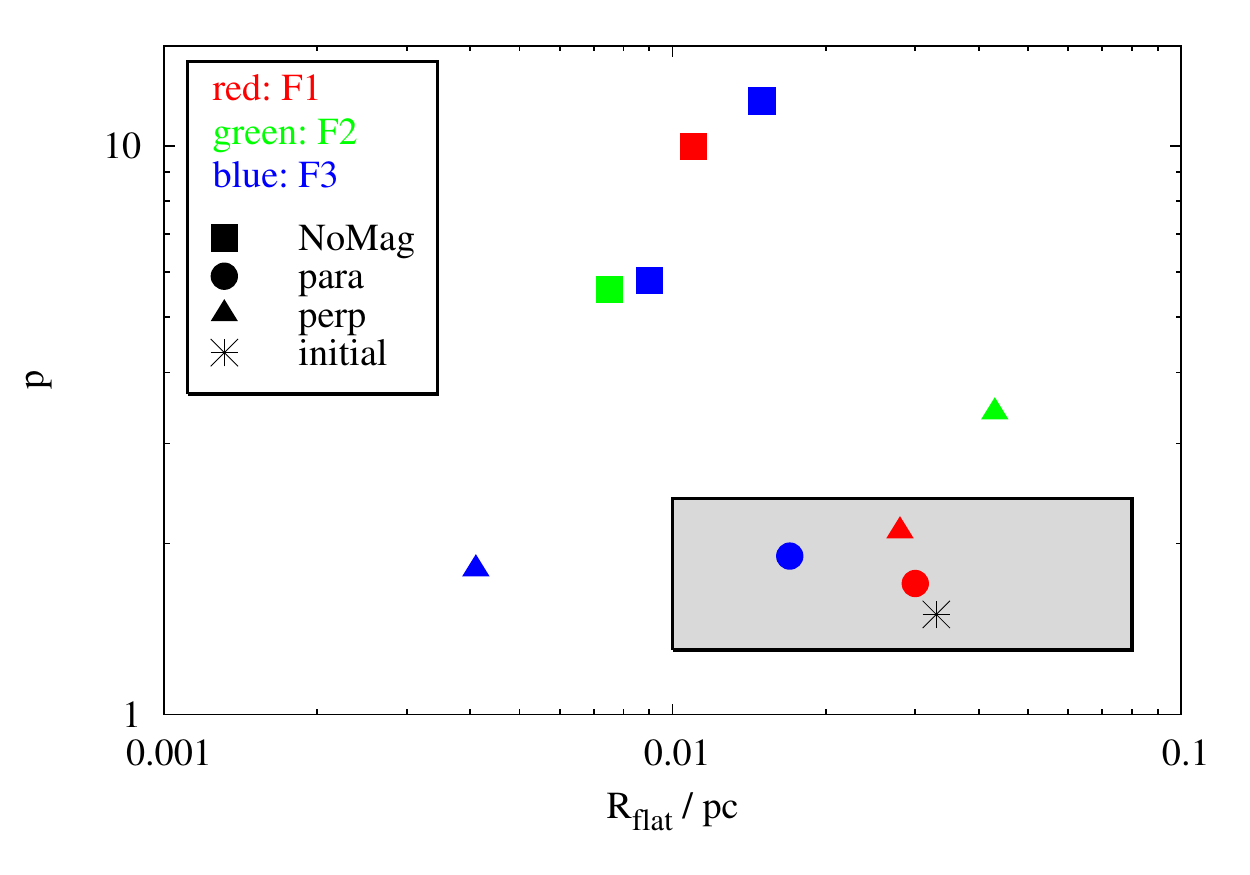}
 \caption{Filament width (here represented with $R_\rmn{flat}$) and exponent $p$ of the density profile at the end of the runs listed in Table~\ref{tab:fits}. Red, green, and blue points correspond to filaments with an initial central density of 1 $\times 10^{-19}$, 1.4 $\times 10^{-19}$, and 3 $\times 10^{-19}$ g cm$^{-3}$, respectively (labelled F1, F2, and F3 as in Table~\ref{tab:models}), squares, circles, and triangles to no magnetic field, a parallel, and a perpendicular magnetic field, respectively. The black asterisk represents the initial parameter setting. The grey-shaded rectangular gives the observational range found by \citet{Arzoumanian11}. Only the runs with a magnetic field parallel to the filament axis are within that range.}
 \label{fig:width}
\end{figure}

We repeat this analysis for the final density profiles of several other runs (see Table~\ref{tab:fits}) and plot $p$ against $R_\rmn{flat}$ at $t_\rmn{end}$ in Fig.~\ref{fig:width}. Only filaments with an initially parallel magnetic field are compatible with observed values indicated by the grey shaded region. For the remaining runs the density profile becomes more narrow and/or declines more steeply. Also for more turbulent initial conditions (run F3\_NoMag\_L2\_M2.5) a rather narrow filament develops, a problem we discuss in more detail in Section~\ref{sec:width}. Interestingly, for run F1\_perp\_L2 the fitting values fall into the observationally determined range as well. However, unlike the other filaments, here no stars are formed. We suppose that in this case the magnetic pressure supporting the filament against collapse in the longitudinal direction as well as the thermal support against radial contraction, i.e. $(M/L)_\rmn{fil} = (M/L)_\rmn{crit}$, are just sufficient to maintain a filament width of about 0.1 pc.

Overall, we find that whether or not the radial density profile of a filament fits the observed values partly depends on the considered time. At late times, all runs but those with a longitudinal magnetic field show a rather narrow filament width and a steep drop-off in disagreement with observational results.

\subsection{Fragmentation spacing}
\label{sec:frag}

In the following, we analyse the mean spacing $\Delta x$ between distinct regions of star formation, where \textit{distinct} implies that we subsume close binary or multiple systems. We focus in our analysis on the filaments with $(M/L)_\rmn{fil} = 3 (M/L)_\rmn{crit}$ and compare $\Delta x$ with the Jeans length $\lambda_\rmn{J}$ (Equation~\ref{eq:jeans}), which for these runs is 0.94 pc. Here, we have assumed a gas temperature of 15 K and a gas density of 3 $\times$ 10$^{-19}$ g cm$^{-3}$. Furthermore, in order to take into account the stabilizing effect of a magnetic field, we can replace $c_s^2$ in equation~\ref{eq:jeans} by $c_\rmn{s}^2 + v_\rmn{A}^2$, where $v_\rmn{A}$ is the Alfv\'en velocity (equation~\ref{eq:Alfven}). Assuming that $v_\rmn{A} \simeq c_\rmn{s}$ (see Section~\ref{sec:IC}), we thus obtain a modified Jeans length, which is larger by a factor of $\sqrt{2}$.

By means of a linear perturbation analysis, \citet{Nagasawa87} showed that the typical distance of fragments in a self-gravitating, isothermal cylinder with a \textit{longitudinal} magnetic field is
\begin{equation}
 \lambda_\rmn{cyl} = 3.52 \times (\pi c_\rmn{s}/G \rho_\rmn{c})^{1/2} \, .
 \label{eq:cylinder}
\end{equation}
Using $T = 15$ K and $\rho_\rmn{c} = 3 \times 10^{-19}$ g cm$^{-3}$, for these ``sausage-type'' instabilities we obtain a typical spacing of $\lambda_\rmn{cyl}$ = 0.33 pc.

\begin{figure}
 \includegraphics[width=\linewidth]{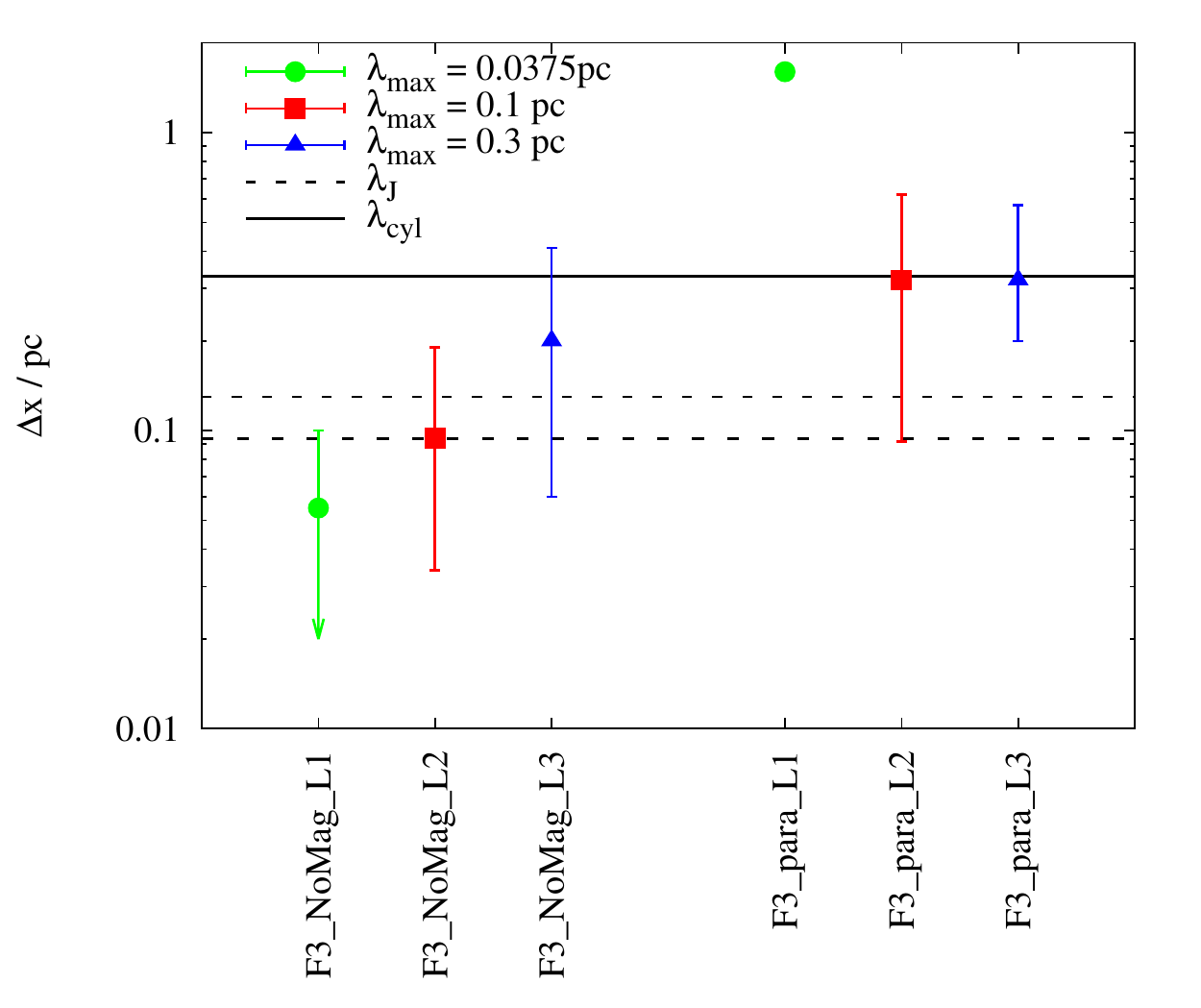}
 \caption{Mean separation $\Delta x$ between the sink particles for the runs F3\_NoMag\_L1, F3\_NoMag\_L2, and F3\_NoMag\_L3 (left side),  and F3\_para\_L1, F3\_para\_L2, and F3\_para\_L3 (right side) after 350 kyr. The error bars show the minimum and maximum separation found in each run, the different colours and symbols distinguish the different turbulent integral scales $\lambda_\rmn{max}$. The two black dashed lines denote the Jeans length $\lambda_\rmn{J}$ excluding (lower line) and including (upper line) the magnetic pressure (see Section~\ref{sec:frag}). The black line shows the spacing $\lambda_\rmn{cyl}$ derived for an isothermal, self-gravitating cylinder. Since in run F3\_para\_L1 only two sink particles have formed, no minimum and maximum spacing is shown.}
 \label{fig:spacing}
\end{figure}

We only consider the filaments with $(M/L)_\rmn{fil} = 3 (M/L)_\rmn{crit}$ and determine the spacing 350 kyr after the start of the simulation. The results are shown in Figure~\ref{fig:spacing} where, in addition to $\Delta x$, the error bar shows the maximum and minimum distance between the distinct sites of star formation.
For run F3\_NoMag\_L2, we identify 18 distinct locations of ongoing star formation. Given the fact that the distances shrink over time due to the global contraction of the filament along its major axis, we normalize the spacing between the fragments to the \textit{initial} length of the filament of 1.6 pc. Hence, for a given number of $N$ star forming sites, we obtain the mean spacing by
\begin{equation}
 \Delta x = \frac{1.6 \, \rmn{pc}}{N - 1} \, .
\end{equation}
For run F3\_NoMag\_L2 this results in a mean spacing of $\Delta x \simeq 0.094$ pc. For run F3\_perp\_L2 (not shown in Figure~\ref{fig:spacing}) we find about 15 different sites of star formation, which results in a mean spacing of about 0.11 pc, which is similar to that of run F3\_NoMag\_L2 despite the larger number of fragments. In both runs $\Delta x$ is in rough agreement with the initial Jeans length of the filament of $\lambda_\rmn{J}$ = 0.094 pc. We point out, however, that this comparison is only approximate since the density of the filament changes during its evolution. Nevertheless, this again demonstrates that a magnetic field perpendicular to the major axis of the filament cannot contribute to the stabilization of a supercritical filament.

For run F3\_para\_L2 the situation is somewhat different: Taking into account the overdense regions around $x \simeq -0.2$ pc and 0.05 pc (see middle panel in Fig.~\ref{fig:timeevol}), which have not yet formed any sink particles, we obtain 8 regions of (future) star formation with a mean spacing $\Delta x$ of about 0.23 pc. Considering only those regions which have actually formed sink particles, the mean distance $\Delta x$ is about 0.32 pc. This is about 2 -- 3 times larger than the distance of the fragments expected from a simple Jeans analysis, which gives $\lambda_\rmn{J} = 0.094$ and 0.13 pc when excluding/including the magnetic field pressure. The measured $\Delta x$, however, is in good agreement with the spacing of the ``sausage-type'' instability of $\lambda_\rmn{cyl}$ = 0.33 pc \citep{Nagasawa87}.

\subsubsection{Impact of the integral scale of turbulence}
\label{sec:intscale}

As shown before, the spacing of fragments can be explained by a Jeans analysis or the instabilities in an isothermal, self-gravitating cylinder \citep{Nagasawa87}. However, one could expect the turbulent motions within the filaments to also influence the fragmentation properties, in particular since the turbulent integral scale $\lambda_\rmn{max} = 0.1$ pc is comparable to the mean distance between the fragments of 0.1 - 0.3 pc reported in Section~\ref{sec:frag}. In order to test this, we have performed additional runs with turbulent integral scales $\lambda_\rmn{max}$ of 0.0375 pc and 0.3 pc both for no magnetic field and a longitudinal field (runs F3\_NoMag\_L1, F3\_NoMag\_L3, F3\_para\_L1, and F3\_para\_L3)

We plot the mean spacing $\Delta x$ of these runs in Figure~\ref{fig:spacing}. For the unmagnetised runs there is a positive correlation between $\Delta x$ and $\lambda_\rmn{max}$, which indicates a strong impact of the turbulent integral scale on $\Delta x$ for unmagnetised, supercritical filaments\footnote{We note that due to the heavy fragmentation taking place in run F3\_NoMag\_L1, it was hard to define a reliable minimum spacing (green arrow).}. Moreover, the poor agreement between $\Delta x$ and the Jeans length for the runs F3\_NoMag\_L1 and F3\_NoMag\_L3 questions whether in the context of filament fragmentation the Jeans analysis is always a useful method.
 
When excluding run F3\_para\_L1, where only two sink particles have formed, for the magnetized runs there is no correlation between $\lambda_\rmn{max}$ and $\Delta x$ recognisable. The spacing is significantly larger than the Jeans length but in excellent agreement with $\lambda_\rmn{cyl}$ (equation~\ref{eq:cylinder}). Moreover, all values of $\Delta x$ are larger than those in the unmagnetised runs, which again emphasizes the stabilizing effect of a longitudinal magnetic field. 

A particular interesting point is the large difference in $\Delta x$ for F3\_NoMag\_L1 and F3\_para\_L1, which we attribute to the same physical origin: The smaller $\lambda_\rmn{max}$, the larger is the fraction of turbulent energy on smaller scales and the faster it gets dissipated \citep{Stone98}. For run F3\_NoMag\_L1 this results in a quick decrease of the turbulent pressure counteracting gravity and in turn enhanced fragmentation. On the other hand, for run F3\_para\_L1, where the magnetic field already accounts for the stabilization of the filament, turbulence appears to decay too quickly to be able to create local overdensities which could lead to additional fragmentation. Hence, the results of F3\_NoMag\_L1 and run F3\_para\_L1 represent the two modes of turbulence in star formation, i.e. \textit{globally} counteracting gravity by means of an additional turbulent pressure and \textit{locally} promoting star formation by creating local overdensities \citep{MacLow04}.

To summarize, our simulations show a high sensitivity of the spacing between distinct sites of star formation on the turbulence field as well as the magnetic field orientation. The simple approach of a Jeans analysis does not match our findings for every case and could even by partly coincidental. Rather, the filaments with a longitudinal magnetic field show a fragmentation behaviour in good agreement with theoretical predictions for a self-gravitating, isothermal cylinder.

\subsection{Effect on turbulence on star formation in filaments}
\label{sec:SF}

Next, we analyse the impact of the initial conditions on the global star formation properties. We show that our initial conditions have only a moderate impact on the global star formation rate although the previously discussed bimodal character of turbulence in star formation is recovered.

In a first step, we consider the time averaged accretion rates $\dot{M}_\rmn{star}$ of all sink particles and the formation times $t_\rmn{form}$ at which the first sink particles are formed (Table~\ref{tab:results}). Apart from run F1\_perp\_L2 the values for $\dot{M}_\rmn{star}$ and $t_\rmn{form}$ in each simulation-subset, i.e. runs with the same initial central density, do not differ too much from each other\footnote{We exclude run F3\_NoMag\_L2\_condensed from the consideration since here already the initial density profile differs.}. In fact, for each subset the total accretion rates lie within about half an order of magnitude and $t_\rmn{form}$ varies by less than a factor of two.

The bimodal effect of turbulence on the star formation process, i.e. locally promoting and globally hampering star formation \citep[see e.g.][for an overview, but see also Section~\ref{sec:intscale}]{MacLow04}, is also recognisable when considering the time when star formation sets in at $t_\rmn{form}$: For the runs with $(M/L)_\rmn{fil} = 3 (M/L)_\rmn{crit}$, $t_\rmn{form}$ seems to decrease with increasing $\lambda_\rmn{max}$ (see Table~\ref{tab:results}). We speculate that since for larger $\lambda_\rmn{max}$, turbulent velocity fluctuations occur also on larger scales, the turbulent energy does not dissipate as fast as in corresponding runs with smaller $\lambda_\rmn{max}$. The turbulent motions rather lead to strong local compressions of the gas which, in turn, results in a faster collapse and an earlier onset of star formation for larger $\lambda_\rmn{max}$. In contrast to that, when increasing the amount of initial turbulence from transonic to supersonic values (F3\_NoMag\_L2 and F3\_NoMag\_L2\_M2.5), it seems that the onset of star formation is somewhat delayed ($t_\rmn{form} = 239$ kyr compared to $t_\rmn{form} = 293$ kyr) due to a higher turbulent pressure. However, the results also show that an enhanced amount of turbulence can by no means reduce the overall accretion rate $\dot{M}_\rmn{star}$ once star formation has set in. Rather, there is a slight increase of about 20\% in $\dot{M}_\rmn{star}$ with respect to $\dot{M}_\rmn{star}$ of the fiducial run F3\_NoMag\_L2. However, since the difference is quite small, we cannot draw reliable conclusions.

\section{Discussion}
\label{sec:discussion}

\subsection{A universal filament width?}
\label{sec:width}

In the previous section we showed that for the runs without any magnetic field as well as a magnetic field perpendicular to the major axis of the filament, the fits of a Plummer-like profile at $t = t_\rmn{end}$ fail to reproduce values comparable to observations (see Table~\ref{tab:fits} and Fig.~\ref{fig:width}), since the filaments become to too narrow. Only filaments with a longitudinal magnetic field reasonably fit the observations for all times. Furthermore, \citet{Fiege00} showed that also a toroidal magnetic field could stabilize the filament against radial collapse. However, observations show that the magnetic field in filaments seems to be preferentially perpendicular to the major axis \citep{Chapman11,Sugitani11,Planck14} although also parallel configurations are observed \citep{Li13,Pillai14,Planck14b}. Under the assumption that there is indeed a universal width for star forming filaments of the order of 0.1 pc as proposed by \citet{Andre13} \citep[or even larger, see][]{Schisano14}, our results raise the question what mechanism allows to retain this width.

As suggested by \citet{Hacar13}, the apparently universal filament width of about 0.1 pc could have its origin in the overlap of a number of parallel, rather narrow filaments, which are too close to each other to be resolved. By means of high-resolution, line emission observations the authors could show that filaments, which appear uniform in dust emission \citep[e.g. Herschel observations by][]{Arzoumanian11}, can split up into a bundle of narrow filaments with widths comparable to those found in our work. Such filamentary substructures were also seen in numerical simulations by \citet{Moeckel14} and \citet{Smith14}. However, these authors follow the entire formation of the filaments in a turbulent environment, which probably sets the stage for an enhanced fragmentation process. In contrast, in our simulations we start from a pre-existing filament, which is most likely the reason why we do not find these sub-filaments. However, our filaments could represent one of these sub-filaments, indicating that such sub-structures would become rather narrow -- as indeed observed.

\subsubsection{Accretion induced turbulence}

\citet{Arzoumanian13} argue that during the evolution of the filament the turbulent support increases or is at least replenished due to accretion driven turbulence \citep{Klessen10,Goldbaum11}, which could stabilize the filament over time. This picture was confirmed by \citet{Heitsch13} by means of semi-analytical calculations. However, as demonstrated in Section~\ref{sec:profiles}, even for an initially supersonic turbulence field (run F3\_NoMag\_L2\_M2.5), the typical width of 0.1 pc cannot be retained for a supercritical filament, even though we allow mass to be accreted onto the filaments.

\begin{figure}
 \includegraphics[width=\linewidth]{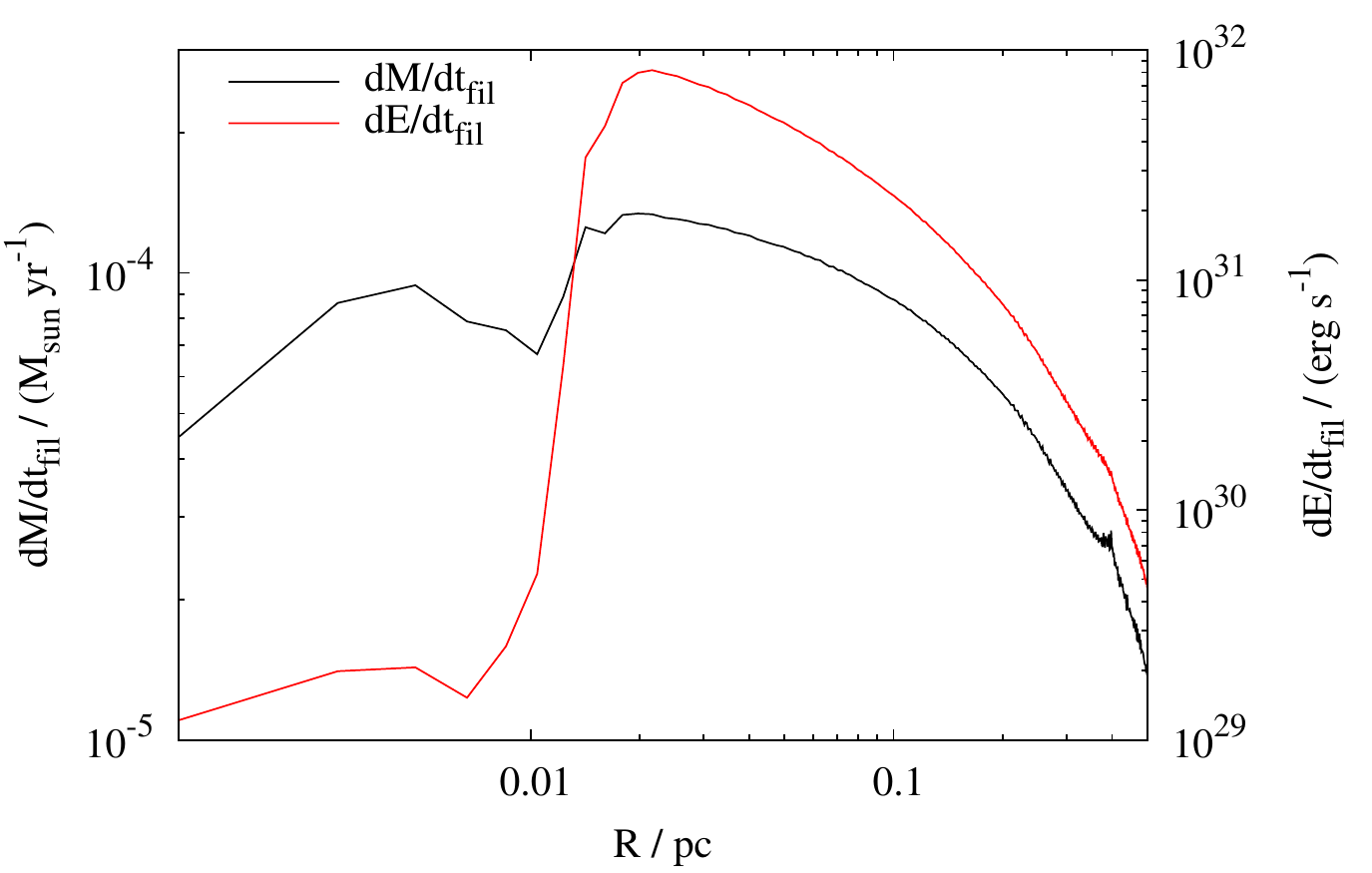}
 \caption{Mass infall rate (black line, left y-axis) and energy infall rate (red line, right y-axis) at the end of run F3\_NoMag\_L2 as a function of the cylindrical radius $R$.}
 \label{fig:accretion}
\end{figure}

In order to investigate this problem in more detail, we analyse the replenishment of turbulent energy in the case of run F3\_NoMag\_L2\footnote{We note that the following estimate gives similar results for other runs as well.}. In Fig.~\ref{fig:accretion} we plot the mass infall rate $\dot{M}_\rmn{fil}(R)$ as a function of the cylindrical radius at the end of the simulation (black line). We obtain $\dot{M}_\rmn{fil}(R) \leq 10^{-4}$ M$_{\sun}$ yr$^{-1}$.
Furthermore, for the accretion driven turbulence model, the inflow rate of kinetic energy is a crucial quantity since it determines the amount of energy, which can -- at least in parts -- be converted into turbulent motions. The energy inflow rate is given by
\begin{equation}
 \left . \frac{dE}{dt} \right |_{\rmn{fil}} = \frac{1}{2} \dot{M}_\rmn{fil}(R) v_\rmn{rad}(R)^2 \, ,
\end{equation}
where $v_\rmn{rad}$ is the radial inflow velocity. We plot $\left . \frac{dE}{dt} \right |_{\rmn{fil}}$ as a function of the radius $R$ in Fig~\ref{fig:accretion} (red line) obtaining values of the order of 10$^{30} - 10^{32}$ erg s$^{-1}$.

As shown by e.g. \citet{MacLow98} and \citet{Stone98}, the turbulent energy is roughly dissipated within one turbulent crossing time:
\begin{equation}
 \left . \frac{dE}{dt} \right |_{\rmn{diss}} = \frac{1}{2} M(R) \sigma^2 / \tau_\rmn{cross} = \frac{1}{2} M(R) \sigma^2 / (R/\sigma) \, ,
\end{equation}
where $\sigma$ is the mean turbulent velocity dispersion, $R$ the radius which we set to 0.05 pc (half of the filament width of 0.1 pc), $M(R) \simeq 15.5$ M$_{\sun}$ the \textit{gas} mass in the filament up to this radius, and $\tau_\rmn{cross} = R/\sigma$ the turbulent crossing time. As stated in \citet{Klessen10}, only a small fraction $\epsilon$ of the infalling energy can be converted into turbulent motions. Hence, in order to find out which amount of turbulence can be sustained by the observed infall rate, we have to solve the following equation for $\sigma$:
\begin{equation}
 \left . \frac{dE}{dt} \right |_{\rmn{diss}} = \frac{1}{2} M(R) \frac{\sigma^3}{R} \stackrel{!}{=} \epsilon \left . \frac{dE}{dt} \right |_{\rmn{fil}}
\end{equation}
Inserting the numeric values given above, we obtain a turbulent velocity dispersion of
\begin{equation}
 \sigma \simeq  0.5 \, \rmn{km \, s^{-1}} \times \epsilon^{1/3} \times  \left ( \frac{\left . \frac{dE}{dt} \right |_{\rmn{fil}}}{10^{31} \rmn{erg \, s}^{-1}} \right )^{1/3} \, .
\end{equation}
Assuming typical values for $\epsilon$ between 10$^{-3}$ and 10$^{-1}$ and using the values for $\left . \frac{dE}{dt} \right |_{\rmn{fil}}$ from Fig.~\ref{fig:accretion}, we obtain values for $\sigma$ of at most 0.5 km s$^{-1}$, i.e. about 2.2 times the sound speed. Hence, with the observed infall rates, at most mildly supersonic turbulent motions can be sustained.  This seems to agree with the results of \citet{Hacar13} who show that (sub-)filaments in general appear to be ``velocity-coherent'' structures. However, with respect to a filament that is supercritical, this turbulence level appears to be relative low and consequently in our simulations the filaments cannot be stabilized against radial collapse.

A possibility to overcome this problem would be a more violent accretion flow, i.e. a larger $\dot{M}_\rmn{fil}$ than the $10^{-5}$ to $10^{-4}$ M$_{\sun}$ yr$^{-1}$ in the simulations presented here. This could be obtained by modelling a denser and gravitationally more unstable molecular cloud environment as done by \citet{Smith14} and \citet{Kirk15}. The authors find filament widths of the order of 0.1 pc and above, although \citet{Smith14} report narrow substructures inside the filaments itself.

To summarize, with the setup used here, supercritical filaments do not reveal a typical width of 0.1 pc at late stages unless a longitudinal magnetic field is present. An initial trans- to supersonic turbulent velocity field as well as the subsequent injection of turbulence by accretion \citep{Klessen10,Arzoumanian13,Heitsch13} are not sufficient to stabilize a filament at a width of 0.1 pc. To overcome this problem, turbulence would have to be replenished in a more efficient way, e.g. by higher inflow velocities than it is the case in our current simulations. Alternatively, filaments could be composed of a bundle of even more narrow, observationally unresolved (sub-)filaments or exhibit an additional longitudinal magnetic field component, which stabilizes the filament along its radial direction.

\subsection{Edge-on vs. centralised collapse}
\label{sec:collapse}

Recently, \citet{Zernickel13} observed a filament which exhibits two major, star forming clumps at either side. On the other hand, observations also show protostellar sources uniformly distributed over the entire filament \citep[e.g.][]{Arzoumanian11,Miettinen12}. In this work we showed that this could have its origin in different masses per unit length compared to the critical value $(M/L)_\rmn{crit}$ (equation~\ref{eq:crit}). However, there are also a number of observations which report the collapse of filaments towards a central massive clump \citep{Myers09,Schneider12,Kirk13,Henshaw14,Peretto14}, which we do not see in the simulations presented so far. A possible mechanism to obtain such a \textit{centralised} collapse mode, is to impose different initial conditions in our simulations. As demonstrated by \citet{Peretto14} there seems to be an increase of the longitudinal, inwards directed velocity towards the common collapse centre. If there exists such an initially converging flow within the filament, this would explain both the velocity gradient as well as the mass accumulating at the convergence point of the flow.

\begin{figure}
 \includegraphics[width=\linewidth]{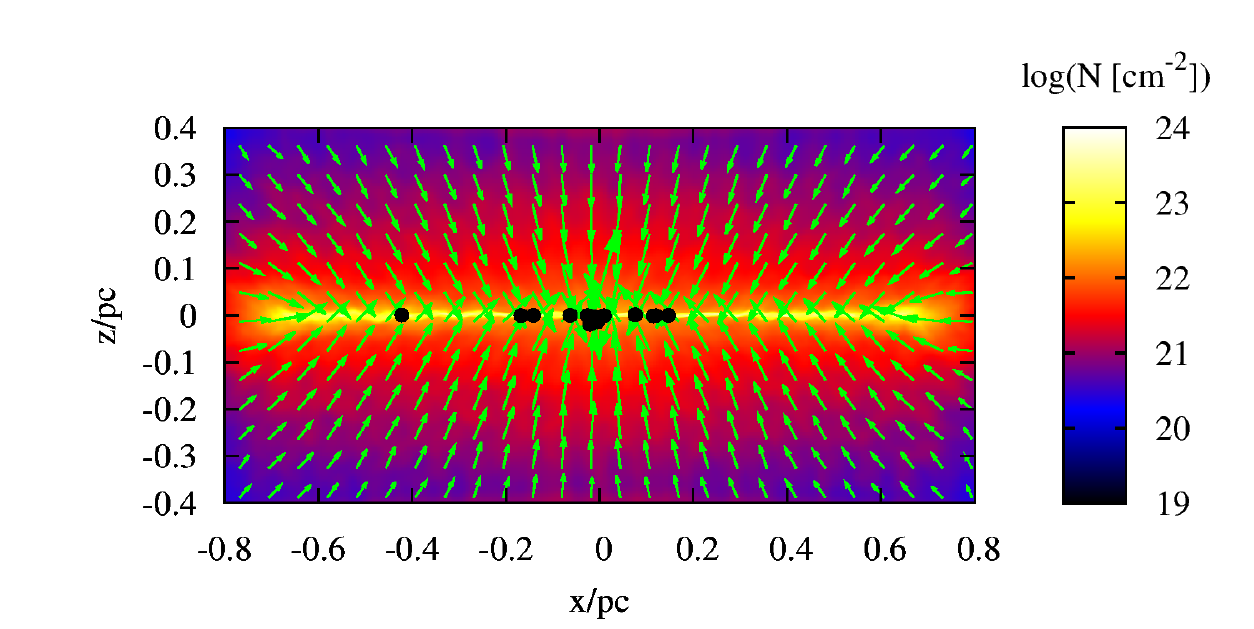}
 \caption{Column density of the filament at the end of run F3\_NoMag\_L2\_condensed. The increased central density results in a global collapse towards the centre.}
 \label{fig:F3_4}
\end{figure}

Another explanation would be that the collapse centre, i.e. the massive clump in the centre of the filament was formed/present already during the formation process of the filament itself. Such a density enhancement could naturally lead to a collapse towards a common centre. In order to test this hypothesis, we consider simulation F3\_NoMag\_L2\_condensed, where the density in the geometrical centre of the filament is enhanced by a factor of three (see equation~\ref{eq:enhance}). In Fig.~\ref{fig:F3_4} we show the column density of the filament at $t_\rmn{end} = 246$ kyr. For this run, star formation already sets in at $t_\rmn{form} = 96$ kyr, which is about three times earlier than in the corresponding run F3\_NoMag\_L2 (Table~\ref{tab:results}).
Even for a moderate density enhancement of a factor of three, a centralised collapse occurs. Fragmentation in the outer parts occurs later, e.g. the sink particle at $x \simeq -0.4$ pc seen in Fig.~\ref{fig:F3_4} forms at $t= 245$ kyr. This agrees with the findings of \citet{Pon11} (see their equation 24) who show that for a perturbation of size $L_1$ and magnitude $\epsilon$ in a cylinder of length 2$L$, the collapse of the perturbation occurs faster than the collapse at the edge for
\begin{equation}
 L_1 < \sqrt{\epsilon} \times L \, .
\end{equation}
Inserting $L_1$ = 0.1 pc, $\epsilon = 2$, and $L = 0.8$ pc as given in our simulation, the above condition is well met, in agreement with the observed centralized collapse mode. Since filaments naturally form in a turbulent environment, it is very likely that density fluctuations will occur along the major axis during their formation. Hence, a centralized collapse mode is easy to obtain for filaments which do not have a uniform density distribution along the major axis.

\subsection{Non-uniform magnetic fields}
\label{sec:caveats}

So far the initial strength of the longitudinal magnetic field in our simulations was constant along the radial direction. In order to account for the observed scaling relation between magnetic fields and the gas density \citep[e.g.][]{Crutcher12} and to test the dependence of our results on the magnetic field configuration in more detail, we perform run F3\_para\_L2\_Brad where the magnetic field strength decreases along the radial direction (see equation~\ref{eq:dropoff}). Compared to run F3\_para\_L2, the total accretion rate $\dot{M}_\rmn{star}$ on all sink particles is about a factor of 2 higher, which is caused by the reduced magnetic pressure. However, the creation time of the first sink particle as well as the number of sinks formed (7 in run F3\_para\_L2 and 8 in run F3\_para\_L2\_Brad) are comparable, which leads to an overall similar fragmentation behaviour.

\begin{figure}
 \includegraphics[width=\linewidth]{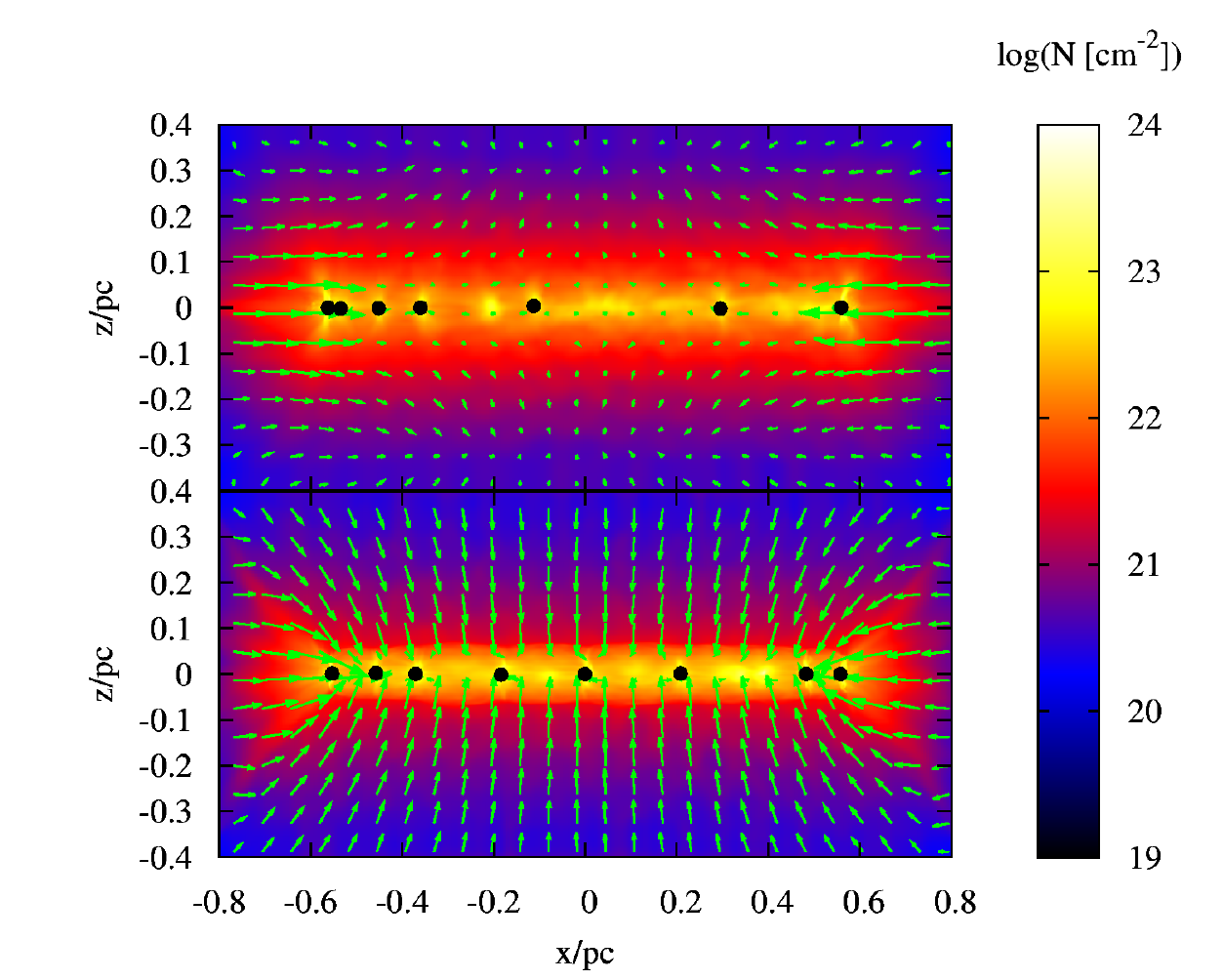}
 \caption{Column density of the filaments of the runs F3\_para\_L2 (top) and F3\_para\_L2\_Brad (bottom) at $t$ = 384 kyr. The initial radial decrease of the magnetic field in run F3\_para\_L2\_Brad results in a more condensed filament.}
 \label{fig:F3_para_4}
\end{figure}

Comparing the column density profile of F3\_para\_L2\_Brad to that of run F3\_para\_L2 at $t = 384$ kyr shows that in run F3\_para\_L2\_Brad the filament is more centrally condensed with a somewhat sharper outer edge (see Fig.~\ref{fig:F3_para_4}). This is also confirmed by fitting the Plummer profile (equation~\ref{eq:fil}) to the radial density profile of the filament, which results in $\rho_\rmn{c} = 20 \times 10^{-19}$ g cm$^{-3}$, $R_\rmn{flat}$ = 0.014 pc, and $p$ = 2.07.
However, the values of $R_\rmn{flat}$ and $p$ only differ by a few 10\% from those of F3\_para\_L2 and by a factor of 2 for $\rho_\rmn{c}$. Since also the star formation properties agree reasonably well, the simplified assumption of a uniform longitudinal magnetic field seems sufficient. 

Although the initial conditions considered are highly idealised, the simulations can serve as a valuable guide for future and more sophisticated simulations including a better treatment of processes like chemistry and radiative transfer. We also point out that, due to the large number of simulations we had to perform for this study, an additional degree of freedom (some degree of disorder for example) would have not been feasible. Moreover, we emphasize that it is difficult to set up a more realistic filament, in particular when magnetic fields are included, which requires the divergence-free condition to be fulfilled. In future work we will therefore investigate the formation and evolution of filaments in molecular clouds in a self-similar fashion by using high resolution, zoom-in simulations of galactic discs \citep{Walch14}. These simulations will also include a proper treatment of the thermodynamical properties of the gas, a fact we so far approximated by an isothermal equation of state. However, given the rather uniform temperatures found in interstellar filaments \citep[e.g.][]{Arzoumanian11}, we consider this as a reasonable approximation.

\section{Conclusions}
\label{sec:conclusions}

In this work we investigate the impact of turbulent motions and different magnetic field configurations on the evolution of interstellar filaments. We also vary the mass per unit length of the filaments ranging from one to three times the critical value. The simulations show a strong impact of turbulence and the magnetic field orientation on the evolution of interstellar filaments. They also provide a valuable guide for additional, more sophisticated simulations.

We consider the radial density profiles of the filaments and compare them with recent observations, which suggest a universal filament width of about 0.1 pc. In general we find that our simulated filaments have a much smaller width and a significantly steeper drop-off of the density at large radii than filaments observed in dust emission \citep[see e.g. the review by][]{Andre13}. In particular, magnetic fields \textit{perpendicular} to the major axis cannot contribute to the stabilization of supercritical filaments ($(M/L)_\rmn{fil} \geq 3 (M/L)_\rmn{crit}$). We note that most of our filaments have rather moderate initial turbulence strengths (Mach numbers around 1), which are at the low side but still in agreement with the observational results. However, even in the presence of supersonic turbulent motions, the filaments quickly become more narrow than $\sim$ 0.1 pc. We also show that accretion driven turbulence can only sustain at most mildly supersonic turbulent motions in our simulations. Hence, this mechanism is not sufficient to support the filaments along their radial direction and to retain a constant width unless a more efficient way is found to convert accretion energy into turbulent motions. In general, these findings fit into the idea that filaments could be composed of bundles of narrow (sub-)filaments \citep{Hacar13}. However, our simulations show that a typical width of 0.1 pc can be retained if the magnetic field is oriented along the major axis of the filament which stabilizes it along the radial direction.

Furthermore, we find that filaments with a mass per unit length equal to the critical one follow an \textit{edge-on} collapse mode with star formation taking place at the outer edges of the filaments in agreement with numerical and (semi-)analytical results \citep{Burkert04,Pon11,Pon12,Clarke15}. No or only little fragmentation is found along the major axes of these filaments. Towards higher masses per unit length, more and more fragmentation is taking place along the entire filament (\textit{uniform} collapse mode). When considering the evolution of a filament with an initial, moderate density enhancement in its centre (factor of three), we could show that it collapses towards this common gravitational centre (\textit{centralized} collapse mode). Our results are thus in good agreement with actual observations of star forming filaments, which discover all three collapse modes, \textit{edge-on}, \textit{uniform}, and \textit{centralized} \citep[e.g.][]{Arzoumanian11,Zernickel13,Peretto14}.

Finally, we find that the distance between fragments along the major axis can only partly be explained by a simple Jeans stability analysis, even when the stabilizing effect of a longitudinal magnetic field is taken into account. For unmagnetised supercritical filaments we could show that turbulent motions have a strong impact on the fragmentation spacing. Hence, we tentatively suggest that the agreement between the fragmentation spacing and the Jeans length could be partly coincidental and that a simple Jeans analysis could be misleading. For filaments with a longitudinal magnetic field, the fragmentation characteristics are reasonably well described by predictions for a self-gravitating cylinder \citep{Nagasawa87}.

\section*{Acknowledgements}

The authors like to thank the anonymous referee for his comments which helped to significantly improve the paper. The authors acknowledge funding by the Bonn-Cologne Graduate School as well as the Deutsche Forschungsgemeinschaft (DFG) via the Sonderforschungsbereich SFB 956 \textit{Conditions and Impact of Star Formation} and the Schwerpunktprogramm SPP 1573 \textit{Physics of the ISM}. The simulations presented here were performed on SuperMUC at the Leibniz Supercomputing Centre in Garching and on JUROPA at the Supercomputing Centre in J\"ulich. The FLASH code was developed partly by the DOE-supported Alliances Center for Astrophysical Thermonuclear Flashes (ASC) at the University of Chicago.

\label{lastpage}

\end{document}